\documentclass[letterpaper]{article} 
\usepackage{aaai2027}  
\usepackage[hyphens]{url}  
\usepackage{graphicx} 
\usepackage{natbib}  
\usepackage{caption} 
\usepackage{algorithm}
\usepackage{algorithmic}

\usepackage{newfloat}
\usepackage{listings}
\DeclareCaptionStyle{ruled}{labelfont=normalfont,labelsep=colon,strut=off} 
\floatstyle{ruled}
\newfloat{listing}{tb}{lst}{}
\floatname{listing}{Listing}

\usepackage{booktabs}

\usepackage{amsmath}
\usepackage{amssymb}
\usepackage{amsthm}
\usepackage{bm}
\usepackage{xcolor}
\usepackage{cleveref}

\crefname{equation}{Eq.}{Eqs.}
\Crefname{equation}{Eq.}{Eqs.}

\title{SQGen: Structured Quantum Image Generation with \\ Latent-Modulated Quantized Tensor Trains}
\author{
    \text{Guang Lin},
    Qibin Zhao
}
\affiliations{
    RIKEN Center for Advanced Intelligence Project (AIP) 
}

\begin{document}

\maketitle

\begin{abstract}
Generating images directly from quantum systems is an attractive but unresolved goal on NISQ hardware.
Existing quantum generators face several coupled obstacles:
barren plateaus that block trainability, expensive quantum circuit preparation, and hardware noise that erodes quantum information with depth. A further difficulty is producing image-scale output without a classical decoder, whose use would otherwise break the end-to-end quantum advantage.
We propose SQGen, a full quantum generator built on a quantized tensor train (QTT) with a latent modulation architecture.
Specifically,
SQGen promotes the QTT bond index of the target pixel distribution to ancilla bond qubits, so that each circuit site operates locally on a bond register plus the two physical qubits that carry the row- and column-bit of one image scale.
We further introduce latent modulation: each re-uploading rotation is factorized at the angle level into a trainable main path plus an additive latent term, reducing to the trainable main path when the latent term is disabled.
During training, we create a differentiable model in the classical system under gate-compatibility constraints, with a torus prior as the latent distribution. After training, every operator maps one-to-one to a native quantum gate, yielding a compact, deployable quantum circuit with no classical decoder in the inference path.
Together, these design choices address the obstacles raised above.
Extensive experiments on image datasets and synthetic data demonstrate that SQGen trains stably, generates images end-to-end from a shallow circuit with no classical decoder, and shows promising feasibility on real quantum hardware.
\end{abstract}

\section{Introduction}
\label{sec:intro}

Quantum generative modeling has emerged as a prominent direction in quantum machine learning, motivated by the prospect that quantum states and circuits offer compact, expressive representations of high-dimensional probability distributions. Parameterized quantum circuits realize this idea through the Born rule, defining distributions that, in certain regimes, are believed to be classically hard to sample \citep{benedetti2019parameterized,coyle2020born,cerezo2021variational}; this very hardness is what a quantum advantage would exploit. The field has since developed several model families: Born machines \citep{liu2018differentiable}, quantum generative adversarial networks \citep{dallaire2018quantum}, and variational quantum generators \citep{romero2021variational}. These early successes establish quantum circuits as expressive distributional priors. However, they have been demonstrated almost entirely on small, low-dimensional targets, leaving the high-dimensional regime, precisely where a sampling advantage would matter, largely untouched.

Scaling quantum generative models to image-scale targets remains substantially harder, and leading approaches fail for distinct reasons. Deep parameterized circuits trained end-to-end can suffer from barren plateaus \citep{mcclean2018barren,cerezo2021cost}, where gradient variance vanishes exponentially with qubit count and circuit depth, causing optimization to stall before useful structure is learned. Amplitude-encoded priors, which map a continuous latent variable to quantum-state amplitudes, incur expensive state-preparation costs \citep{cacioppo2023quantum,kolle2024quantum}, and the deep preparation circuits they require can further erode the encoded information.

Underlying these methods is a shared failure: image-scale quantum generation demands deep, complex circuits, yet such depth is what NISQ hardware cannot run, as device noise accumulates with every additional layer \citep{wang2021noise}. Hybrid CNN-quantum pipelines \citep{huang2021experimental,chen2025quantum,islam2026quantum} instead produce high-quality image samples with a shallow quantum component, but place the bulk of their modeling capacity in a classical decoder, so any quantum-advantage claim attaches to the classical component that is not itself quantum.

To address these challenges, we propose two architectural ideas with a fully classical training pipeline. First, a QTT bond skeleton is made physical inside the circuit: each bond index is promoted to a small ancilla register whose size scales logarithmically with the bond dimension, and each site applies a local unitary on that register together with the two physical qubits encoding one image scale. The model output is the bond-marginal Born distribution, recovered by sampling all qubits and discarding the ancillas.
Second, we introduce latent modulation as a refinement that factorizes each rotation into a trainable main path plus an additive latent term, an exact angle-level decomposition guaranteed by abelian rotation composition.
Finally, training is fully classical: we optimize a differentiable QTT model with exact gradients, avoiding both barren plateaus and the sampling overhead of expectation-style objectives. After training, the gate-compatible parameters export one-to-one to a native-gate quantum circuit, with no classical decoder in inference.

We benchmark SQGen on MNIST datasets, with controlled synthetic targets and a real-hardware study, across generation quality (KL and $L_1$), diversity, and circuit complexity.
Across our evaluated settings, SQGen trains stably and attains the lowest per-sample KL and $L_1$ on all ten digit classes while keeping diversity non-zero. The ablation exposes a synergy between the two components: the bond skeleton alone reduces to a static class prototype, latent modulation alone may be harmful without the bond, and only their combination delivers both sharp class structure and latent-conditional variation. Notably, the trained circuit generates end-to-end with no classical decoder, and shows promising feasibility when executed on real quantum hardware.
In summary, our main contributions are as follows:

\begin{itemize}
    \item We propose SQGen, a structured quantum image generator built on two ideas: a QTT bond skeleton that learns multi-scale image features efficiently with only a few qubits; and latent modulation, with an exact angle-level latent injection that conditions the output on the latent and improves sample diversity.
    \item We develop a fully classical training pipeline, where a differentiable QTT model is optimized with exact gradients under a torus prior and the reconstruction loss using a discardable encoder, enabling one-to-one parameter export to a native-gate quantum circuit without a classical decoder at inference.
    \item We empirically demonstrate the effectiveness of SQGen through extensive experiments, showing that it trains free of barren plateaus, achieves advanced generation performance, and is promising for real quantum hardware.
\end{itemize}

\section{Background}
\label{sec:background}

\paragraph{Quantum circuits and Born machines.}
\label{sec:bg_circuit}

A quantum circuit on $n$ qubits acts on a state $|\psi\rangle \in (\mathbb{C}^2)^{\otimes n}$ by a sequence of unitary gates; we use Dirac notation, so $|i\rangle$ denotes a column basis vector indexed by $i \in \{0,1\}^n$ and $\langle i| = |i\rangle^\dagger$ is its conjugate transpose. We restrict to the native gate set $\{R_y(\theta), R_z(\theta), \mathrm{CNOT}\}$, with Pauli rotations $R_a(\theta) = \exp(-i \theta\, \sigma^a / 2)$ for axis $a \in \{y, z\}$. Applying a circuit $U$ to the initial state $|0\rangle^{\otimes n}$ and measuring in the computational basis defines the Born distribution, $p(i) \;=\; \big| \langle i \mid U \mid 0^{\otimes n}\rangle \big|^2, i \in \{0,1\}^n$.
A parameterized quantum circuit $U_\theta$ has gate angles that depend on trainable parameters $\theta$, and optionally on a continuous input $z$; a quantum circuit Born machine (QCBM) \citep{liu2018differentiable} fits a target distribution by minimizing a divergence against the Born distribution of $U_\theta$. Trainability of deep parameterized circuits is constrained by the barren plateau phenomenon \citep{mcclean2018barren}, in which gradient variance shrinks exponentially with qubit count and circuit depth.

\paragraph{Quantized tensor train.}
\label{sec:bg_qtt}

A tensor train (TT) decomposes a function of $d$ discrete indices as a chain of small cores connected by bond indices,
\begin{equation}
T(x_1, \dots, x_d) \;=\; \sum_{b_1, \dots, b_{d-1}} \prod_{k=1}^{d} A_k\!\left[b_{k-1}, x_k, b_k\right],
\label{eq:bg_tt}
\end{equation}
with cores $A_k \in \mathbb{R}^{\chi_{k-1} \times |x_k| \times \chi_k}$ and boundaries $b_0 = b_d = 1$. The bond dimension $\chi$ controls expressivity: at $\chi = 1$ the TT factorizes across indices, and larger $\chi$ covers progressively richer correlations.
A quantized tensor train (QTT) \citep{khoromskij2011d} first maps a one-dimensional index $X \in \{0, \dots, 2^L - 1\}$ to its coarse-to-fine binary bits and then applies a TT over those bits, yielding a scale-by-scale representation.
For a two-dimensional image, the row and column coordinates each contribute $L$ bits; grouping them by scale yields $L$ levels of two bits each, one row-bit and one column-bit per scale, i.e., factorized as $(2_1 \times 2_2 \times \cdots \times 2_L) \times (2_1 \times 2_2 \times \cdots \times 2_L)$.
This binary, scale-by-scale structure is a natural fit for quantum encoding: each scale is carried by two qubits, and the QTT cores supply the local unitaries that act on them.

\section{Method}
\label{sec:method}

\begin{figure}[t]
\centering
\includegraphics[width=\linewidth]{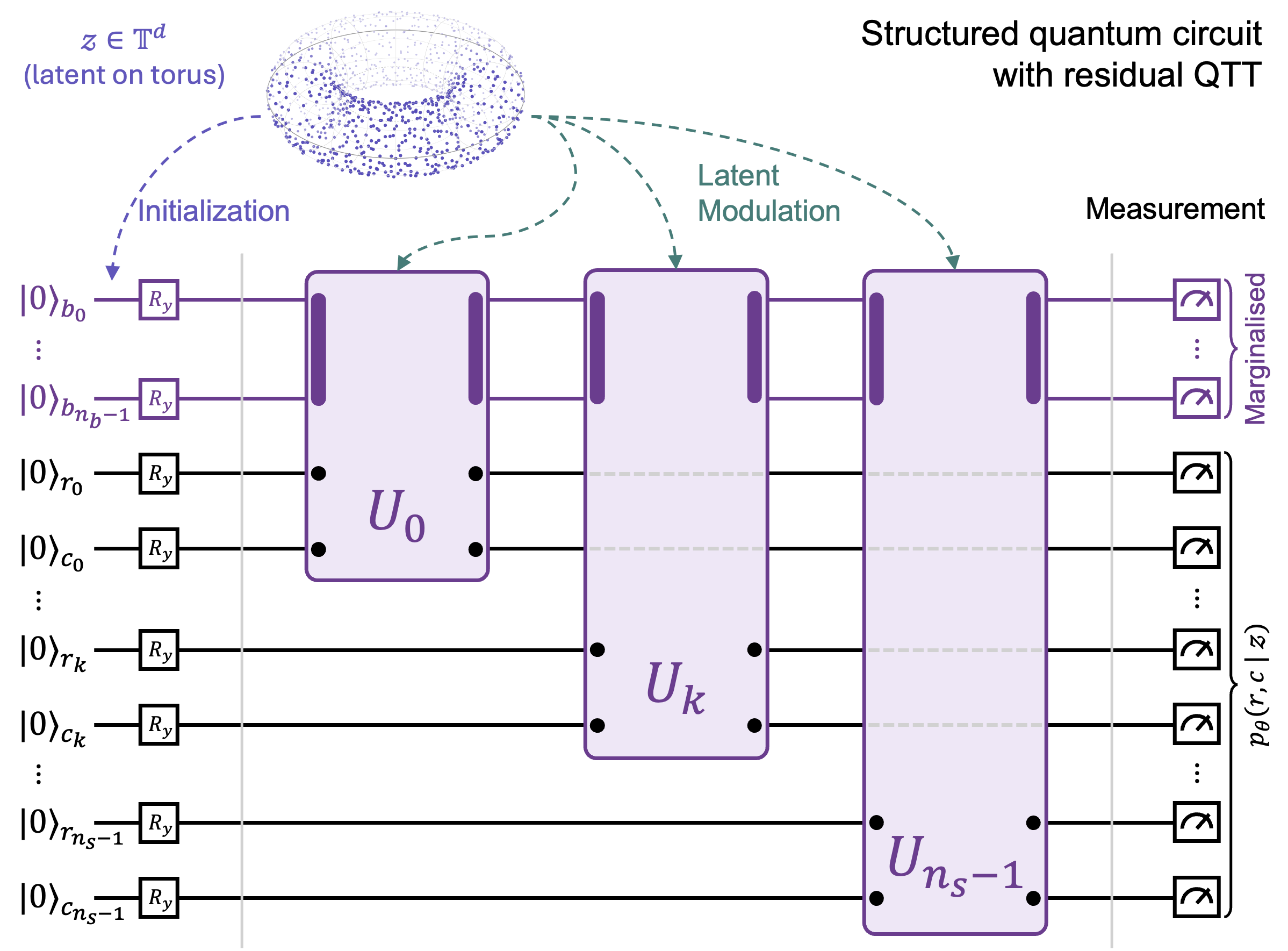}
\caption{The illustration of the SQGen with latent-modulated quantized tensor train skeleton.}
\label{fig:residual_qtt}
\end{figure}

SQGen is a Born-machine image generator whose circuit is designed around a coarse-to-fine image representation and trained entirely on the classical side. We build it on a latent-modulated quantized tensor train (QTT) skeleton (\Cref{fig:residual_qtt}), a multi-scale bond structure whose rotations are conditioned on the latent. Every block is constrained to be unitary, so the trained model maps one-to-one onto a native-gate circuit, a natural fit for quantum deployment.

\subsection{Image mapping and notation}
\label{sec:encoding}

An image of side $s = 2^{n_s}$ is represented as a probability distribution over $2 n_s$ physical qubits, where $n_s$ is the number of multiresolution scales. We use a Morton (Z-order) layout: qubit $q = 2k$ stores the row bit $r_k$ and qubit $q = 2k+1$ stores the column bit $c_k$, with bits indexed from the coarsest scale ($k = 0$) to the finest ($k = n_s - 1$). Thus, each qubit pair $(q_{2k}, q_{2k+1})$ encodes one spatial scale and successive pairs refine the location from coarse to fine, as detailed in \Cref{app:morton}. The target distribution is defined as $q(r,c) \propto x(r,c)$, where $x$ is the image.
A Born machine with parameters $\theta$ and latent $z$ outputs
\begin{equation}
p_\theta(i \mid z) \;=\; \left| \langle i \mid U_\theta(z) \mid 0^{\otimes n}\rangle \right|^2 .
\label{eq:born}
\end{equation}

\subsection{Bond-augmented QTT skeleton}
\label{sec:qtt_skeleton}

Classically, a QTT decomposition of $p(r,c)$ writes
\begin{equation}
\begin{aligned}
&p(r_0, c_0, \dots, r_{n_s-1}, c_{n_s-1}) \\ &\quad \;=\; \sum_{b_1, \dots, b_{n_s-1}} \prod_{k=0}^{n_s-1} A_k\!\left[b_k,\, (r_k, c_k),\, b_{k+1}\right],
\label{eq:qtt_classical}
\end{aligned}
\end{equation}
where the bond indices $b_k \in \{1,\dots,\chi\}$ glue successive sites and $b_0 = b_{n_s} = 1$. Each site $A_k$ is a tensor of shape $(\chi, 4, \chi)$, with the index of size $4$ enumerating the joint state $(r_k, c_k) \in \{0,1\}^2$ of the row and column bits at scale $k$.

\paragraph{Promotion to qubits.}
We make this skeleton physical by allocating $n_b = \log_2 \chi$ ancilla bond qubits and replacing each $A_k$ by a unitary block acting on the bond register $\mathcal{H}_b = (\mathbb{C}^2)^{\otimes n_b}$ together with the two physical qubits encoding $(r_k, c_k)$.
The full unitary $U_\theta(z) = U_{n_s-1} \cdots U_1 U_0$ is applied to the all-$|+\rangle$ state $|{+}^{\otimes (n_b + n)}\rangle = \prod_q R_y(\pi/2)\, |0^{\otimes (n_b + n)}\rangle$, and the model output is the bond-marginal Born distribution:
\begin{equation}
p_\theta(r, c \mid z) \;=\; \sum_{b \in \{0,1\}^{n_b}} \left| \langle b, r, c \mid U_\theta(z) \mid {+}^{\otimes (n_b + n)}\rangle \right|^2 .
\label{eq:bond_marginal}
\end{equation}
The image distribution is recovered directly by measuring all qubits in the computational basis and discarding the bond bits, without any classical decoder or post-processing network in the inference path.

\paragraph{Local site block.}
Each site $k$ acts on the local register $\mathcal{L}_k = \{q_b : b < n_b\} \cup \{q_{r_k}, q_{c_k}\}$, the $n_b$ bond qubits together with the two physical qubits of scale $k$. A site is built from $L_{\text{site}}$ sub-layers; in each sub-layer $\ell$, every local qubit $q \in \mathcal{L}_k$ receives two single-qubit rotations,
\begin{equation}
R_y\!\left(\alpha^y_{k,\ell,q}(z)\right)\, R_z\!\left(\alpha^z_{k,\ell,q}(z)\right),
\label{eq:site_sublayer}
\end{equation}
followed by a fixed CNOT entangler $E_{k,\ell}$, as detailed in \Cref{app:gate_counts}. The effective angles $\alpha^a_{k,\ell,q}(z)$ combine trainable parameters with the latent modulation defined in \Cref{sec:residual}.

\paragraph{Expressivity hierarchy.}
The bond rank $\chi$ provides an explicit capacity knob for cross-scale correlations. When $\chi = 1$, no bond qubits are allocated, and the QTT skeleton reduces to a scale-separable form,
\begin{equation}
p_\theta(r, c \mid z) \;=\; \prod_{k=0}^{n_s-1} p_\theta^{(k)}(r_k, c_k \mid z),
\label{eq:chi_1_factorization}
\end{equation}
so correlations between different spatial scales cannot be mediated by the bond register.
For larger $\chi$, the bond register provides a higher-dimensional channel through which information can pass across QTT sites, allowing the circuit to represent richer cross-scale dependencies.
Thus, $\chi$ acts as an explicit expressivity knob: increasing it enlarges the structured family of distributions accessible to the circuit, yielding a nested capacity hierarchy.

\subsection{Latent modulation (LM)}
\label{sec:residual}

\paragraph{Angle-level modulation.}
We inject the latent $z$ by refining data re-uploading \citep{perez2020data,schuld2021effect} at the level of rotation angles. For a single qubit $q$, axis $a \in \{y, z\}$, and sub-layer $\ell$, the standard re-uploading rotation $R_a(\omega^a_{\ell,q}\!\cdot z + \beta^a_{\ell,q})$ composed with the trainable rotation $R_a(\theta^a_{\ell,q})$ collapses exactly into a single rotation,
\begin{equation}
\begin{aligned}
&R_a\!\left(\theta^a_{\ell,q}\right) R_a\!\left(\omega^a_{\ell,q}\!\cdot z + \beta^a_{\ell,q}\right) \\ &\quad = R_a\!\Big( \underbrace{\theta^a_{\ell,q}}_{\text{main path}} + \underbrace{\beta^a_{\ell,q} + \omega^a_{\ell,q}\!\cdot z}_{\text{latent term}} \Big),
\end{aligned}
\label{eq:angle_residual_identity}
\end{equation}
because rotations about a common axis form an abelian one-parameter subgroup. The effective angle at $(\ell, q, a)$ is
\begin{equation}
\alpha^{a,\,\text{eff}}_{\ell,q}(z) \;=\; \theta^a_{\ell,q} \;+\; \beta^a_{\ell,q} \;+\; \omega^a_{\ell,q}\!\cdot z .
\label{eq:effective_angle}
\end{equation}

Here, $\theta^a_{\ell,q}$ defines the trainable main path, $\omega^a_{\ell,q}\!\cdot z$ is the additive latent term, and $\beta^a_{\ell,q}$ is a learnable offset. Restoring the site index $k$, these effective angles are the $\alpha^a_{k,\ell,q}(z)$ used in \Cref{eq:site_sublayer}. We call this construction latent modulation.

\paragraph{Relation to data re-uploading.}
Data re-uploading places the data-injection rotation next to each trainable rotation. Latent modulation instead combines the two into a single explicit rotation: by the same-axis identity~(\Cref{eq:angle_residual_identity}), the trainable and data rotations are exactly equivalent to one rotation whose angle is a trainable main path, a learnable offset, and an additive latent-dependent term. This explicit additive form is what we build on, and it brings two benefits. First, the rewriting is exact, since it follows from an algebraic identity rather than an approximation; moreover, disabling the modulation weights leaves the trainable rotation block intact up to the learnable offset. Second, in deployment, the same-axis rotations can be fused into a single native rotation, reducing the executed single-qubit rotation count. Both formulations realize the same band-limited function class; what differs is the explicitness of the conditioning and the deployed gate cost, as discussed in \Cref{app:residual_vs_reuploading}.

\subsection{Classical training, quantum deployment}
\label{sec:training_deployment}

Unlike previous works, we first train the QTT model in a classical system and then exported one-to-one to a native-gate quantum circuit for deployment. We describe the latent prior and encoder, the reconstruction objective, and the training and deployment pipeline in turn.

\paragraph{Latent prior and encoder.}
Rather than using the isotropic Gaussian prior common in classical generative models \citep{kingma2013auto}, we place the latent variable $z$ on the torus $\mathbb{T}^d$ and sample it from a scrambled Sobol quasi-Monte Carlo (QMC) sequence. The torus matches the $2\pi$ periodicity of the rotation-angle encoding, so the prior support coincides with the domain on which $z$ acts as a circuit angle. The Sobol point set further replaces independent random samples with a low-discrepancy set, giving more uniform coverage of the latent domain and a lower-variance prior estimate at fixed batch size; details are given in \Cref{app:torus_prior}. To amortize training over images, a small CNN-based encoder $E_\phi: x \mapsto z \in \mathbb{T}^d$ maps each image to a torus latent, which is matched to the prior by a squared maximum mean discrepancy (MMD) penalty
\begin{equation}
\mathcal{L}_{\text{prior}} = \mathrm{MMD}^2\!\left(\Phi(E_\phi[\mathcal{B}]),\Phi(\tilde{\mathcal{Z}})\right),
\label{eq:mmd_prior}
\end{equation}
where $E_\phi[\mathcal{B}]=\{E_\phi(x):x\in\mathcal{B}\}$, $\mathcal{B}$ is a minibatch of images, $\tilde{\mathcal{Z}}$ is a set of Sobol prior points of the same size, and $\Phi(z)=[\cos z,\sin z]$ respects the $2\pi$-periodic geometry of the torus. The encoder is used only during training and discarded at deployment.

\paragraph{Reconstruction objective.}
The encoder and QTT model are trained by per-image reconstruction: for each image $x$, the Born distribution at $z = E_\phi(x)$ is matched to the corresponding target pixel distribution $q_x$, rather than matching the data distribution at the population level. The reconstruction objective combines a Kullback-Leibler (KL) divergence and an $L_1$ distance:
\begin{equation}
\mathcal{L}_{\text{rec}}(p_\theta,q_x) = \lambda_{\text{KL}}\, \mathrm{KL}(q_x\|p_\theta) + \lambda_{L_1}\, \|p_\theta-q_x\|_1 ,
\label{eq:rec_loss}
\end{equation}
where $\lambda_{\text{KL}}$ and $\lambda_{L_1}$ are hyperparameters.

\begin{figure}[t]
\centering
\includegraphics[width=\linewidth]{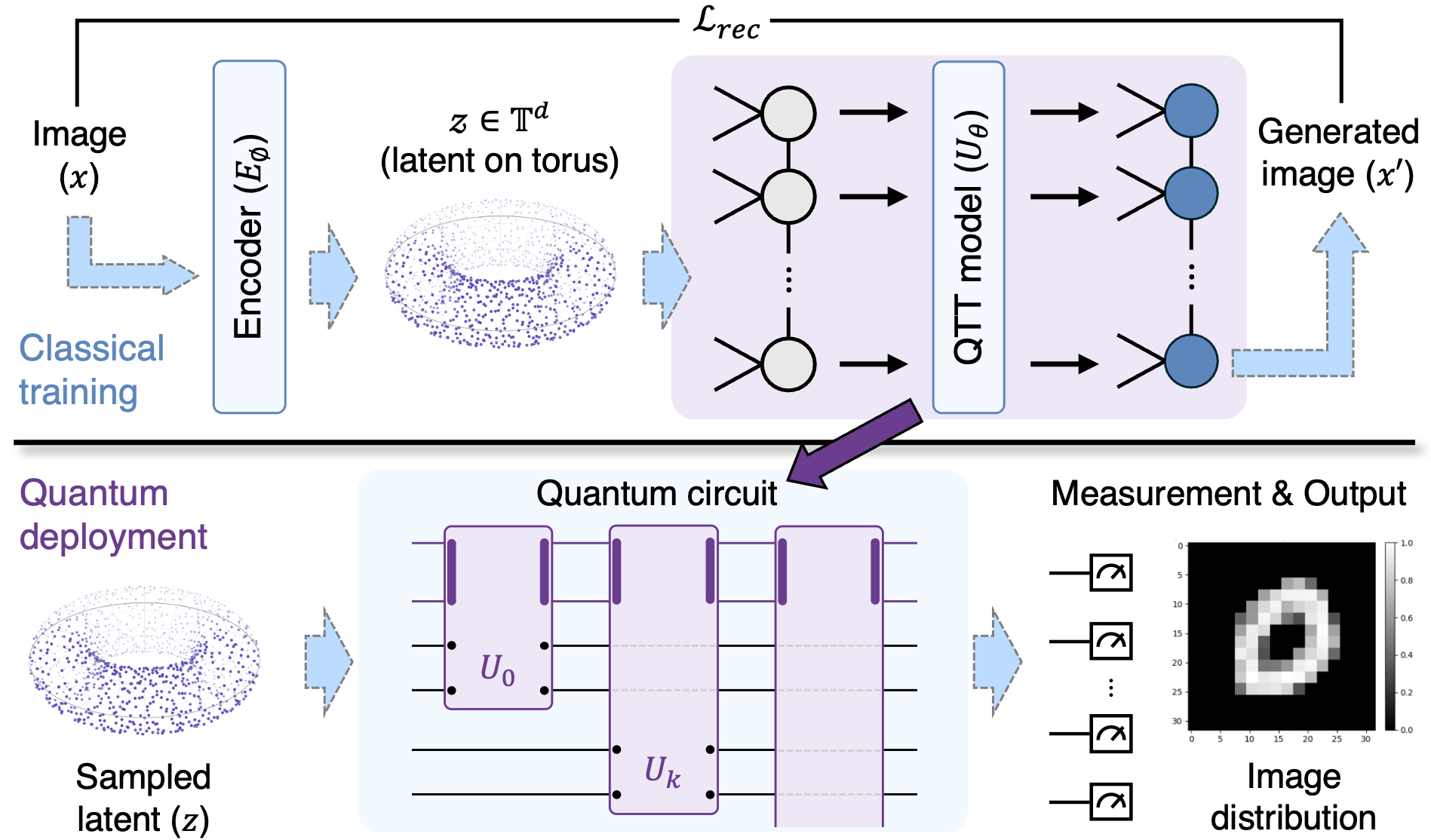}
\caption{The illustration of classical training and quantum deployment pipelines.}
\label{fig:training_deployment}
\end{figure}

\paragraph{Training and deployment pipeline.}
As shown in \Cref{fig:training_deployment}, training is performed entirely on the classical system. The encoder $E_\phi$ and the gate-compatible QTT model $U_\theta$ are optimized jointly with AdamW \citep{loshchilov2017decoupled}, where $p_\theta(\cdot \mid z)$ is computed by differentiable tensor contractions and exact gradients are backpropagated through both the encoder and the parameterized QTT tensors. This avoids finite-shot sampling overhead and removes quantum-circuit simulation from the training loop. After training, the encoder is discarded and the learned QTT parameters are exported one-to-one to a native-gate quantum circuit. At deployment, the input of quantum circuits is sampled directly from the torus prior on $\mathbb{T}^d$, the circuit is measured in the computational basis, and the image distribution is obtained by marginalizing out the bond ancillas.

\section{Related Work}
\label{sec:related}

\paragraph{Quantum generative models.}
Many quantum generative models follow the variational quantum algorithm (VQA) paradigm \citep{cerezo2021variational}, in which a parameterized circuit is trained by a classical optimizer, with model families distinguished mainly by their training objective. Quantum circuit Born machines \citep{liu2018differentiable} define implicit generative models through the Born rule, with sampling believed to be classically hard for certain circuit families \citep{coyle2020born}. Quantum generative adversarial networks \citep{dallaire2018quantum,lloyd2018quantum} replace explicit-divergence training with an adversarial discriminator, while variational quantum generators \citep{romero2021variational} target continuous outputs through amplitude-based constructions. Existing demonstrations in this line are largely concentrated on low-dimensional targets. SQGen differs by targeting image-scale, decoder-free conditional generation with a gate-compatible Born machine trained using exact classical gradients.

\paragraph{Hybrid classical-quantum generators.}
A common route to visually plausible image samples combines a small quantum component with a classical decoder \citep{huang2021experimental,chen2025quantum,islam2026quantum}. Such hybrid pipelines can improve sample quality, but much of the representational capacity resides on the classical side, and the decoder remains in the inference path, which weakens any end-to-end quantum-advantage claim. SQGen takes a different design choice: a classical encoder is used only for training and is discarded at deployment. The deployed model is a native-gate quantum circuit whose output pixel distribution is read directly from computational-basis measurements with the bond ancillas marginalized.

\paragraph{Tensor networks and quantum circuits.}
Tensor networks provide compact low-rank representations for structured high-dimensional data, and several works compile matrix product state (MPS) or tensor train (TT) representations into quantum circuits by mapping tensor cores to sequences of unitary blocks \citep{han2018unsupervised,ran2020encoding,lin2026structured}. In classical machine learning, tensor-network priors have also been used directly for supervised learning \citep{stoudenmire2016supervised} and unsupervised generative modeling \citep{meiburg2025generative}, showing that low-rank decompositions can provide useful inductive bias for structured data. SQGen connects these directions by making the QTT bond structure physical inside the quantum circuit. The QTT skeleton fixes the circuit topology, while expressivity is supplied by trainable native-gate rotations rather than by post-hoc compilation of an unconstrained tensor network.

\begin{figure*}[t]
\centering
\includegraphics[width=\linewidth]{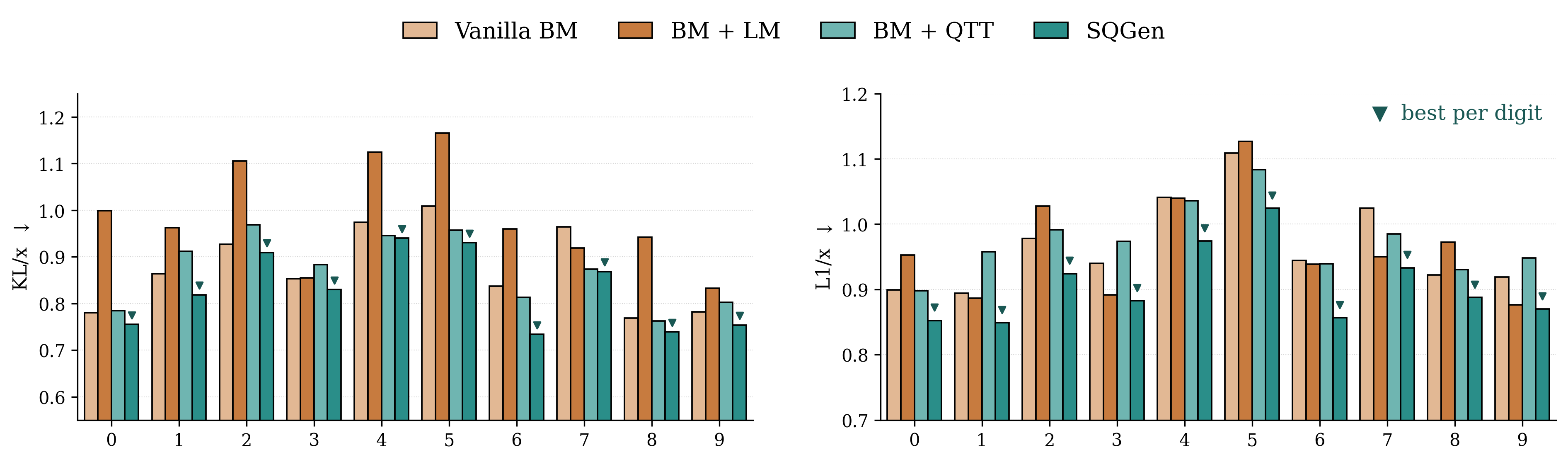}
\caption{MNIST image generation on per digit. Side-by-side bars compare the four configurations on $\mathrm{KL/x}$ (left) and $\mathrm{L1/x}$ (right); $\nabla$ marks the winner per digit.}
\label{fig:per_digit_kl}
\end{figure*}

\section{Experiments}
\label{sec:experiments}

We evaluate SQGen on several image distribution generation tasks and the results demonstrate that SQGen outperforms the traditional methods and ablations across diverse metrics. In addition, we perform SQGen on the real quantum hardware, showing the feasibility of our method in practice.

\subsection{Setup}
\label{sec:setup}

\paragraph{Datasets.}
We evaluate SQGen on MNIST \citep{lecun1998mnist}, Fashion-MNIST \citep{xiao2017fashion} datasets, and a suite of synthetic controlled targets, including stripes and Gaussian blobs. Each image is zero-padded to $2^{n_s} \times 2^{n_s} $ pixel grid matching the physical-qubit budget, and treated as a per-class conditional pixel distribution.

\paragraph{Evaluation metrics.}
We report three families of metrics: per-sample reconstruction quality, measured by the KL divergence and $L_1$ distance between each generated and target pixel distribution; class-mean fidelity; and sample diversity. We complement these with visualizations of generated samples. In addition, we also conduct the experiments on real quantum hardware, showing the feasibility in practice.

\paragraph{Implementation details.}
The proposed method is trained end-to-end on a differentiable model implemented in PyTorch \citep{paszke2019pytorch}, then the trained model is exported to a quantum circuit in Qiskit \citep{javadi2024quantum}. More details are reported in \Cref{app:setup_details}.

\subsection{Image generation on MNIST datasets}
\label{sec:main_results}

\begin{figure}[t]
\centering
\includegraphics[width=\linewidth]{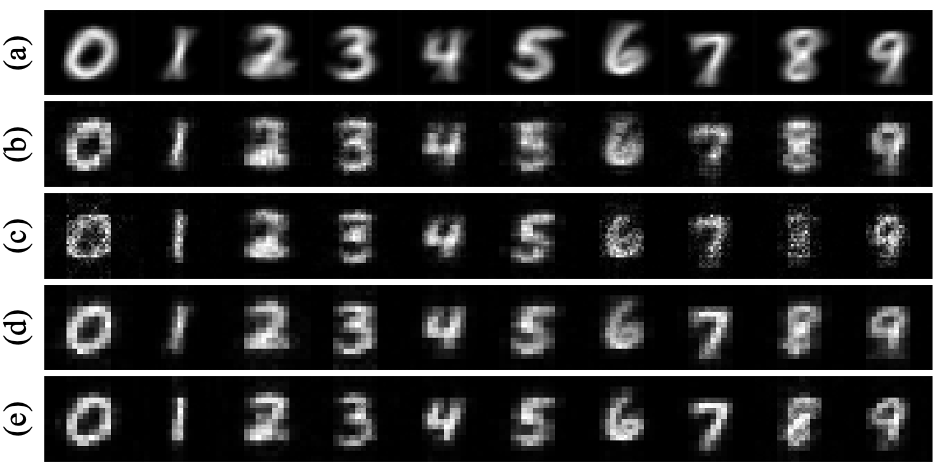}
\caption{The generated MNIST samples across the four configurations. (a) empirical class mean (target prototype); (b) Vanilla BM; (c) BM+LM; (d) BM+QTT; (e) SQGen.}
\label{fig:synergy_grid}
\end{figure}

\Cref{tab:mnist_aggregated} reports the four configurations alongside two trivial baselines (a class-mean predictor and a uniform distribution). The per-sample reconstruction errors $\mathrm{KL/x}$ and $\mathrm{L1/x}$ average the KL divergence and $L_1$ distance between each generated distribution and its target over test images. The class-mean scores $\mathrm{KL}_{\text{c}}$ and $\mathrm{L1}_{\text{c}}$ compare the per-class average output against the empirical class mean; and diversity (Div.) is the mean pairwise $L_1$ distance between a class's samples.
Among the four configurations, SQGen attains the lowest $\mathrm{KL/x}$ and $\mathrm{L1/x}$ and the highest diversity. Its per-sample reconstruction is competitive with the static class-mean predictor, the zero-diversity reference that marks the reconstruction floor: SQGen is only slightly behind on $\mathrm{KL/x}$ and in fact lower on $\mathrm{L1/x}$, while, unlike that predictor, producing genuinely varied samples. The per-digit view in \Cref{fig:per_digit_kl} confirms this is not an averaging artifact, with SQGen winning both $\mathrm{KL/x}$ and $\mathrm{L1/x}$ on all ten digits; the per-digit numbers are tabulated in \Cref{app:mnist_per_digit}. \Cref{fig:synergy_grid} shows the comparison visually, each column giving one digit class generated by all configurations. In addition, we further conduct the experiments on Fashion-MNIST, as reported in \Cref{tab:fashion_aggregated}.

\paragraph{Ablation analysis.}
The four configuration comparison shows that the QTT bond skeleton and latent modulation (LM) are individually insufficient and pay off in combination.
With the bond but no modulation, BM+QTT collapses to a sharp but $z$-invariant prototype whose diversity is exactly zero, so it produces no image-level variation.
With modulation but no bond, BM+LM injects $z$-dependent variation but loses per-sample fidelity, giving the worst $\mathrm{KL/x}$ of the four. \Cref{fig:synergy_grid} shows the effect directly: modulation alone (b$\to$c) yields visibly noisy samples, the bond alone (b$\to$d) recovers a faithful but static prototype, and their combination (e) reproduces the empirical class mean (a) with genuine per-image variation in \Cref{tab:mnist_aggregated}.
The two mechanisms are thus complementary rather than substitutable, and only SQGen secures class-faithful structure and latent-conditional variation at once. The experiments on Fashion-MNIST dataset also show similar results, reported in \Cref{tab:fashion_aggregated} and \Cref{tab:fashion_per_class}.

\begin{table}[t]
\centering
\setlength{\tabcolsep}{4pt}
{\small
\begin{tabular}{lrrrrr}
\toprule
Method
  & $\mathrm{KL/x}$ $\downarrow$
  & $\mathrm{L1/x}$ $\downarrow$
  & $\mathrm{KL}_{\text{c}}$
  & $\mathrm{L1}_{\text{c}}$
  & Div. $\uparrow$ \\
\midrule
\multicolumn{6}{l}{\emph{Four configurations}} \\
Vanilla BM           & 0.876 & 0.967 & 0.141 & 0.299 & 0 \\
                        & $\pm 0.088$ & $\pm 0.070$ & $\pm 0.034$ & $\pm 0.046$ & / \\
BM + LM                 & 0.986 & 0.966 & 0.282 & 0.455 & 0.184 \\
                        & $\pm 0.113$ & $\pm 0.079$ & $\pm 0.073$ & $\pm 0.063$ & $\pm 0.082$ \\
BM + QTT                & 0.870 & 0.974 & 0.129 & 0.290 & 0 \\
                        & $\pm 0.076$ & $\pm 0.054$ & $\pm 0.042$ & $\pm 0.055$ & / \\
SQGen  & \textbf{0.828} & \textbf{0.905} & 0.171 & 0.348 & \textbf{0.214} \\
                        & $\pm 0.081$    & $\pm 0.058$    & $\pm 0.066$ & $\pm 0.059$ & $\pm 0.175$ \\
\midrule
Class mean              & 0.753 & 0.918 & 0     & 0     & 0 \\
Uniform                 & 2.121 & 1.719 & 1.362 & 1.396   & 0 \\
\bottomrule
\end{tabular}
}
\caption{MNIST image generation, aggregated over 10 digits (mean $\pm$ std). The class-mean baseline trivially achieves $\mathrm{KL}_{\text{c}} = 0$ and is reported only as a lower-bound reference.}
\label{tab:mnist_aggregated}
\end{table}

\subsection{Synthetic controlled targets}
\label{sec:synthetic}

\begin{figure}[t]
\centering
\includegraphics[width=\linewidth]{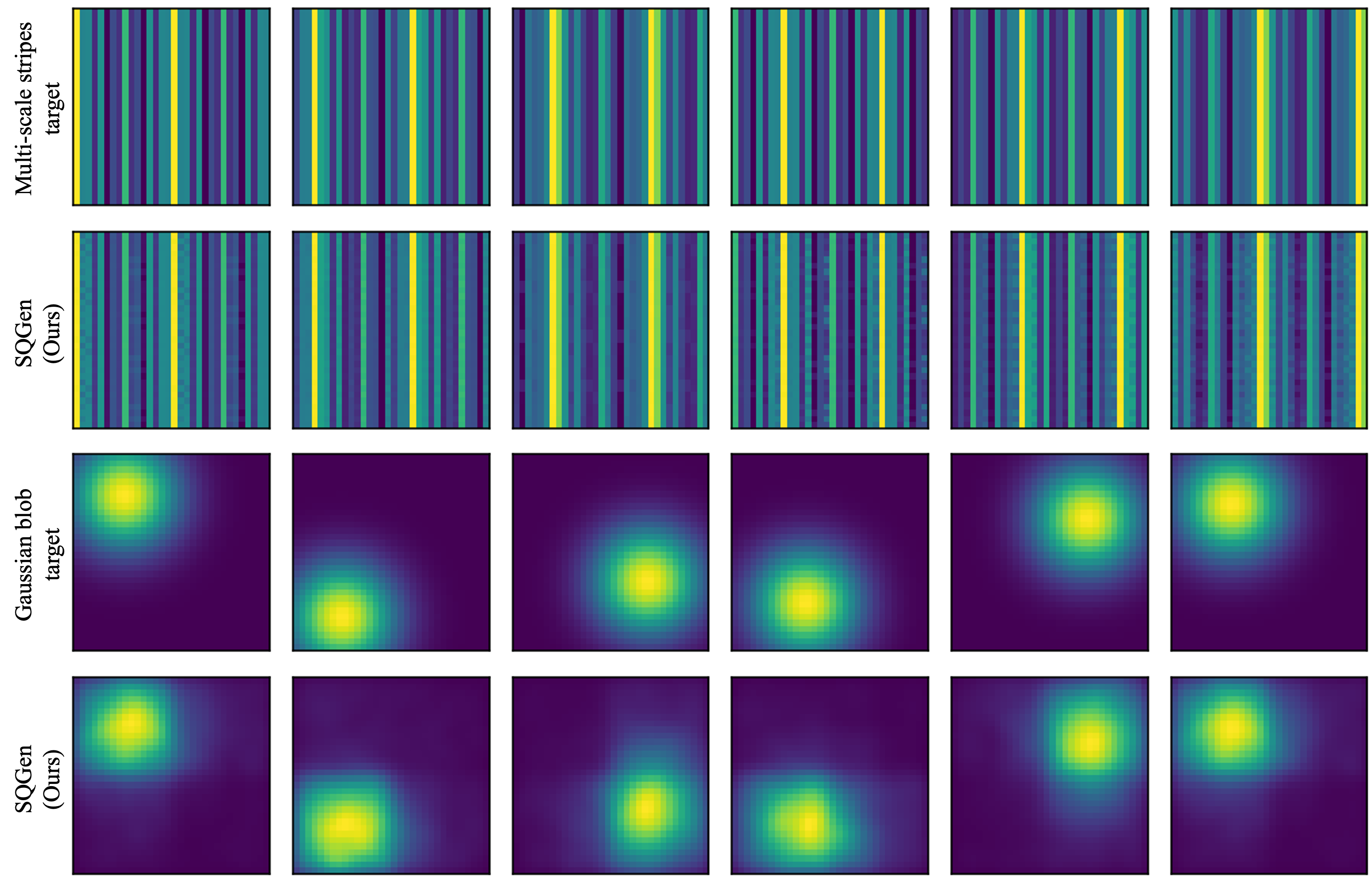}
\caption{Synthetic controlled targets and SQGen outputs. Top (rows 1, 2): a multi-scale cosine sum on the column axis; Bottom (rows 3, 4): $2$-D Gaussian blobs on $z \in \mathbb{T}^2$.}
\label{fig:synthetic}
\end{figure}

We further evaluate SQGen on two controlled targets with known structure (\Cref{fig:synthetic}). These tasks are not intended as standalone image-generation benchmarks; rather, they provide visually interpretable tests of two behaviors needed by SQGen: representing multi-scale spatial patterns and producing latent-dependent changes in the output distribution.

\paragraph{Multi-scale stripes.}
The first target is a $z$-invariant multi-scale stripe pattern formed by combining column-wise cosine components at several spatial frequencies. This target provides a simple visual test of whether the model can represent coarse and fine spatial variations simultaneously. As shown in \Cref{fig:synthetic}, SQGen closely reproduces the stripe structure.

\paragraph{Continuous-position Gaussian blob.}
The second target is a two-dimensional Gaussian blob whose center changes with the latent variable $z\in\mathbb{T}^2$. This task tests whether the model output varies consistently with a continuous latent input, rather than collapsing to a fixed template. As shown in \Cref{fig:synthetic} (bottom), the generated blobs move with the target locations across Sobol latent draws. These controlled examples provide qualitative evidence for the intended behavior of the architecture; the main quantitative image-generation results are reported in \Cref{sec:main_results}.

\subsection{Bond-rank scaling}
\label{sec:ablations}

We sweep $\chi \in \{2, 4, 8, 16\}$ with all other hyperparameters fixed at the defaults of \Cref{tab:mnist_aggregated}, reporting MNIST digit~$0$ in \Cref{tab:bondrank} and the average over all ten digits in \Cref{tab:bondrank_agg}.

\paragraph{Bond-rank sensitivity.}
The rank of tensor networks (TN) is the main capacity parameter in TN representation. A larger rank allows the TN model to transmit richer information across scales and thus represent more detailed spatial structure, but it also increases the bond size and the associated cost. As shown in \Cref{tab:bondrank}, the generation quality of SQGen improves as $\chi$ increases from 2 to 4, while the benefit largely saturates at higher ranks. These results are consistent with classical TN literature.

\paragraph{Default bond-rank selection.}
We further repeat the bond-rank sweep across all ten MNIST digits and report the averaged results in \Cref{tab:bondrank_agg}. The aggregate trend shows that larger $\chi$ does not lead to uniformly better performance: reconstruction improves from small ranks to moderate ranks, but the gains saturate and become less stable at larger ranks. In addition, diversity is highest at small $\chi$ and decreases as $\chi$ grows, since a larger bond rank tightens the model's representation of the class mean at the cost of per-sample variation. Among the tested values, $\chi=8$ achieves the best average reconstruction while maintaining non-zero sample diversity, making it a practical default for the headline configuration. This choice also balances the additional expressivity provided by the bond register against the increased circuit cost introduced by larger ranks.

\begin{table}[t]
\centering
\setlength{\tabcolsep}{8pt}
\begin{tabular}{lccccc}
\toprule
Method & $\chi$ & $\mathrm{KL/x}$ & $\mathrm{L1/x}$ & $\mathrm{KL}_{\text{c}}$ & $\mathrm{L1}_{\text{c}}$ \\
\midrule
BM+QTT     & 2  & 0.813 & 0.925 & 0.224 & 0.416 \\
BM+QTT     & 4  & 0.787 & 0.905 & 0.198 & 0.386 \\
BM+QTT     & 8  & 0.785 & 0.898 & 0.196 & 0.360 \\
BM+QTT     & 16 & 0.799 & 0.892 & 0.210 & 0.360 \\
\midrule
SQGen & 2  & 1.158 & 1.031 & 0.368 & 0.549 \\
SQGen & 4  & 0.754 & 0.853 & 0.230 & 0.411 \\
SQGen & 8  & 0.755 & 0.852 & 0.166 & 0.341 \\
SQGen & 16 & 0.733 & 0.841 & 0.207 & 0.384 \\
\bottomrule
\end{tabular}
\caption{Bond-rank sweep on MNIST dataset.}
\label{tab:bondrank}
\end{table}

\subsection{Training stability}
\label{sec:barren}
\begin{figure}[t]
\centering
\includegraphics[width=\linewidth]{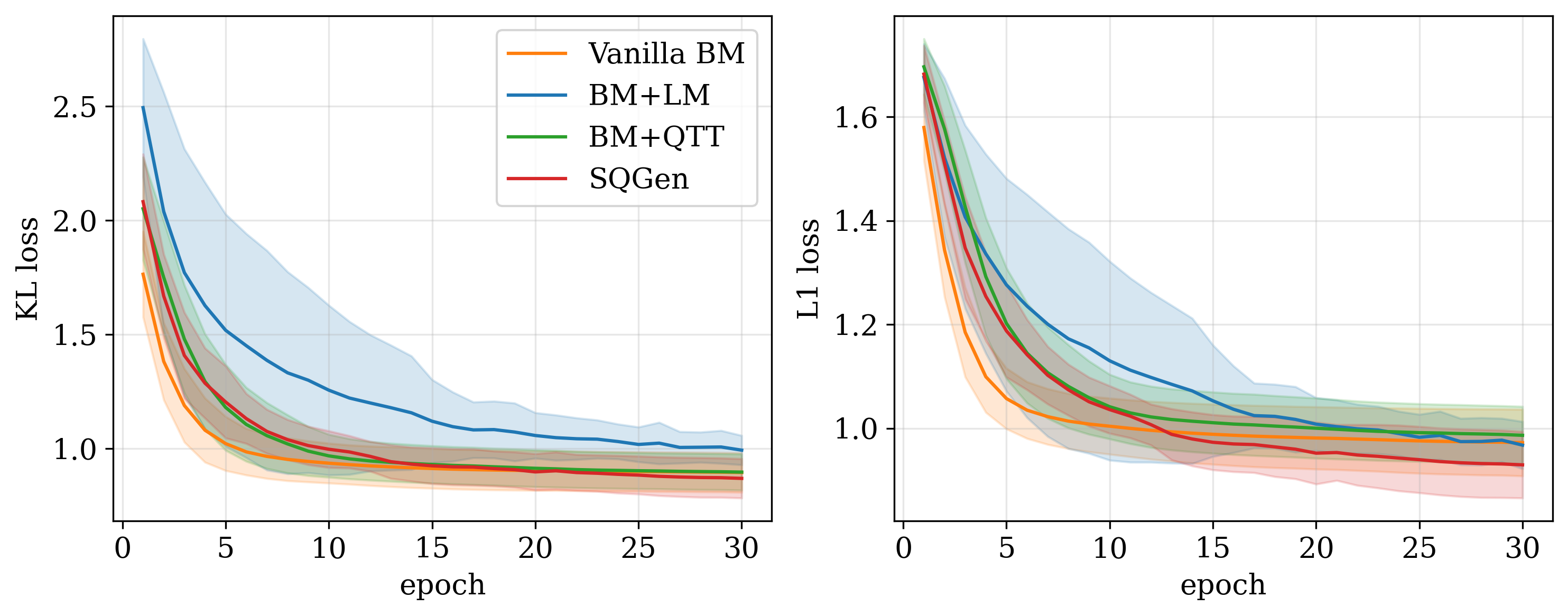}
\caption{Training loss curves (mean $\pm$ std over the ten MNIST digits): KL loss (left) and $L_1$ loss (right) versus epoch for the four configurations.}
\label{fig:training_curves}
\end{figure}

One motivation for training SQGen on the classical side is to avoid the practical trainability barriers of deep variational quantum circuits, where barren plateaus can flatten gradients and stall optimization \citep{mcclean2018barren,cerezo2021cost}. In our experiments, SQGen does not exhibit empirical optimization collapse. \Cref{fig:training_curves} reports the KL and $L_1$ training losses averaged over the ten MNIST digits for the four configurations. All curves decrease steadily from initialization, without an extended flat phase at the beginning of training, and SQGen reaches the lowest final KL and $L_1$ losses.
This behavior is consistent with the intended role of the classical training pipeline. SQGen is optimized as a differentiable QTT model with exact backpropagated gradients, rather than through shot-based estimates of quantum expectation values. As a result, the training loop avoids the shot noise and measurement overhead that can make barren-plateau regimes practically inaccessible in classical variational training.

\subsection{Noise robustness}
\label{sec:noise}

We evaluate the deployed circuits under a controlled depolarizing noise setting. Specifically, after every two-qubit CNOT gate, we apply a depolarizing channel with rate $p$; the channel is defined in \Cref{app:depolarizing_noise}. To isolate the effect of the injected noise from shot-based sampling, we compute exact output probabilities using the Aer simulator on digit 0. At $p=0$, the curves recover the corresponding noiseless reconstruction behavior, so the degradation as $p$ increases reflects the sensitivity of the circuit to the noise.

\begin{figure}[htbp]
\centering
\includegraphics[width=\linewidth]{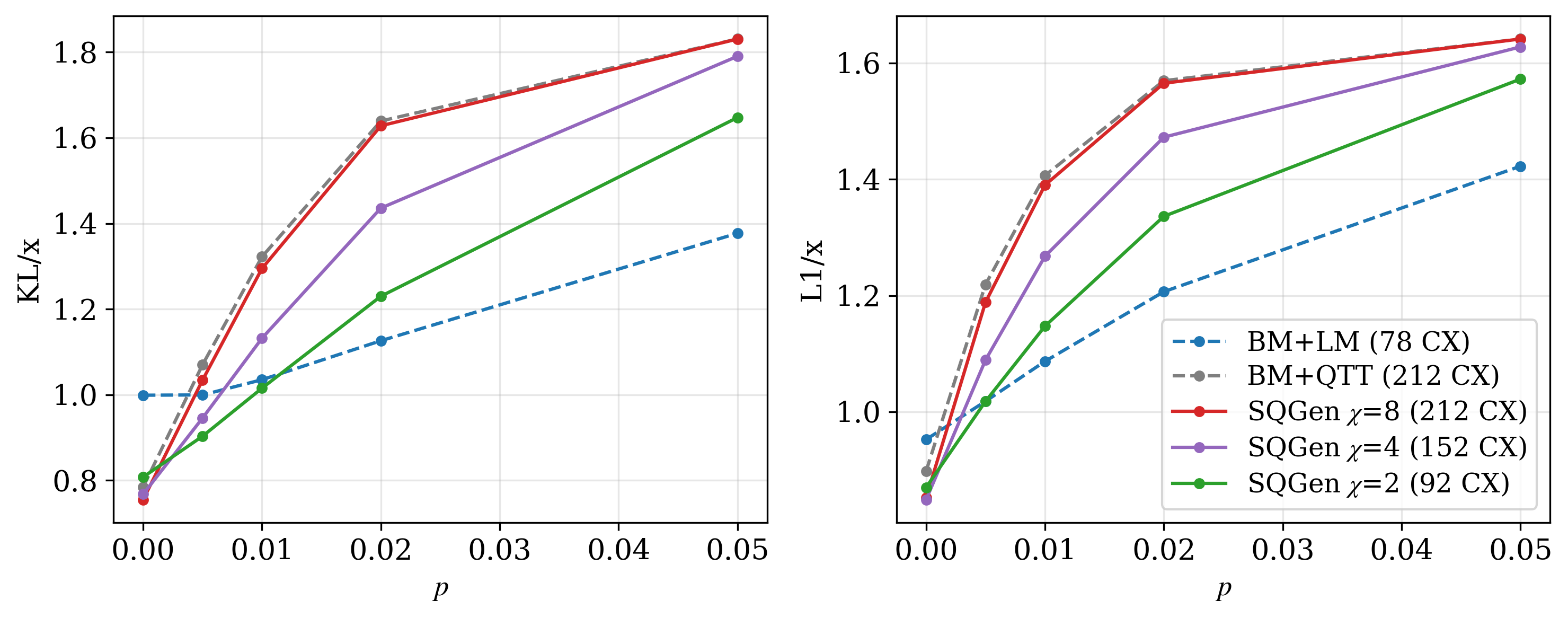}
\caption{Noise robustness under exact Aer simulation. A depolarizing channel is applied after every CNOT gate, and we report per-sample $\mathrm{KL/x}$ (left) and $\mathrm{L1/x}$ (right) as functions of the noise rate $p$.}
\label{fig:noise}
\end{figure}

All methods degrade as the depolarizing rate increases. The degradation is more pronounced for circuits with larger CNOT counts, since the noise channel is applied after every two-qubit gate. In particular, the default SQGen configuration with $\chi=8$ has the largest bond register among the tested variants and therefore accumulates more injected noise under this model. This illustrates a basic robustness trade-off in the structured circuit: increasing the bond rank improves representational capacity in the noiseless setting, but also increases the number of noisy two-qubit operations. The bond rank $\chi$ provides a direct way to tune this trade-off. Reducing $\chi$ decreases the number of bond qubits and CNOT gates, producing flatter degradation curves in \Cref{fig:noise}. The $\chi=2$ and $\chi=4$ variants are therefore less sensitive to the injected depolarizing noise than the default $\chi=8$ circuit, although they also provide less cross-scale capacity. Thus, the appropriate bond rank depends on the desired balance between representational capacity and robustness under the assumed noise level.

Overall, this experiment should be interpreted as a controlled noise-sensitivity study. In practice, current quantum hardware operates at low per-gate error rates, reported in \Cref{app:hardware}. A thorough study of the strong-noise regime is valuable in its own right, but it is beyond the scope of this paper and we leave it to future work.

\subsection{Real quantum hardware deployment}
\label{sec:real_hardware}

\begin{figure}[htbp]
\centering
\includegraphics[width=\linewidth]{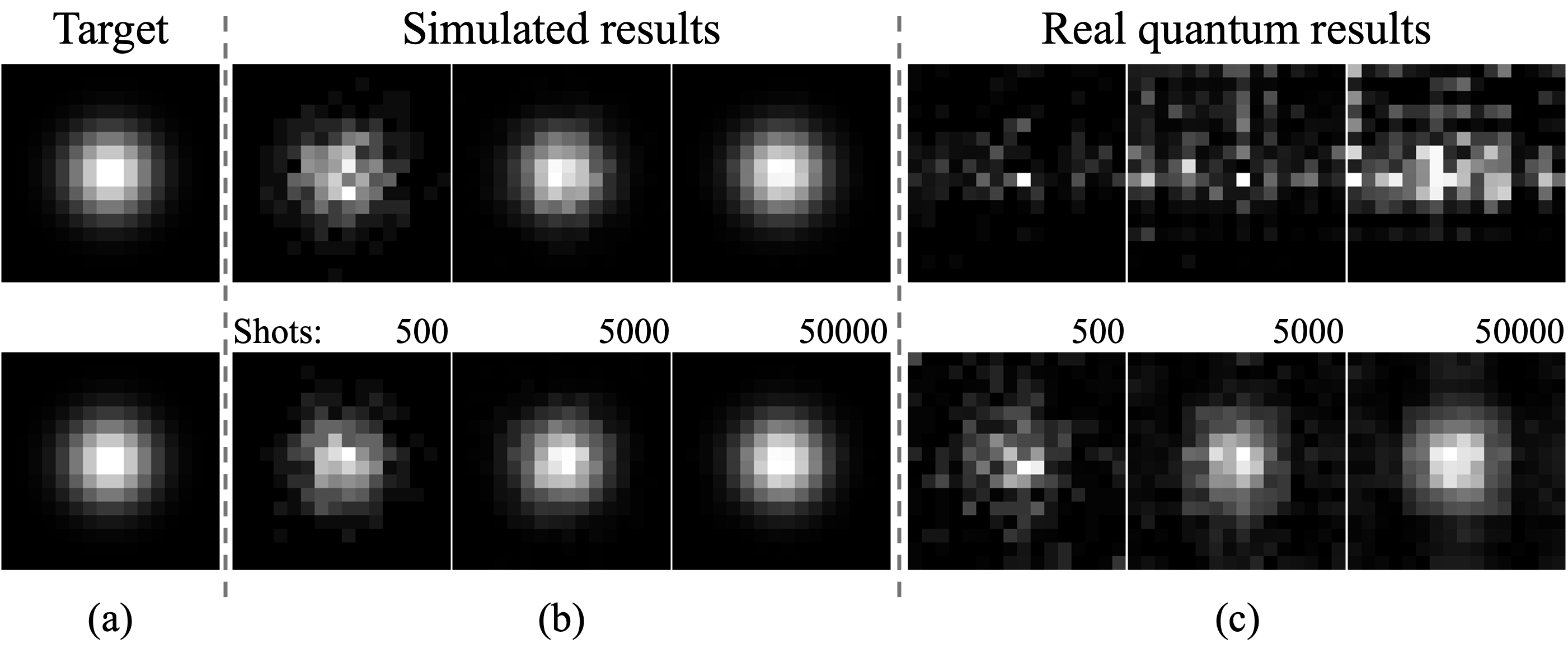}
\caption{Visualization on simulation and real hardware.}
\label{fig:real}
\end{figure}

We further deploy the trained SQGen circuit on real quantum hardware (\texttt{ibm\_kawasaki} on the IBM Quantum Platform; device details in \Cref{app:hardware}) and compare it with an amplitude-encoding (AE) baseline, which generates the image through an expensive state-preparation circuit. \Cref{fig:real} compares the target images, noiseless simulation results, and real quantum hardware outputs. In the simulation, both SQGen and the AE-based baseline recover the target blob structure reasonably well, indicating that the corresponding ideal circuits encode the desired distributions. However, on real quantum hardware, their behavior differs substantially. SQGen is still able to generate images that closely resemble the target, whereas the AE-based baseline is strongly degraded and no longer retains the useful structure. This contrast suggests that SQGen is more compatible with NISQ hardware, providing promising feasibility in practice.
\section{Conclusion}
\label{sec:conclusion}

We propose SQGen, a structured quantum image generator built on three ideas: a quantized tensor train (QTT) bond skeleton made physical inside the circuit; latent modulation (LM) as an exact angle-level factorization into a trainable main path and an additive latent term; and a classical-training, quantum-deployment pipeline that optimizes a differentiable gate-compatible QTT model before exporting it one-to-one to a native-gate circuit. Across extensive experiments on image datasets and synthetic data, we empirically demonstrated that SQGen trains stably, generates images end-to-end from a shallow circuit with no classical decoder, and shows promising feasibility on real quantum hardware.

\bibliography{aaai2027}

\newpage
\appendix

\section{Pixel-basis mapping details}
\label{app:morton}

We use a Morton (Z-order) layout to align image pixels with the QTT site ordering. For an image of side $2^{n_s}$, the row and column coordinates are written in binary as
\begin{equation}
r=(r_0,\ldots,r_{n_s-1}),
\quad
c=(c_0,\ldots,c_{n_s-1}),
\end{equation}
where $r_0,c_0$ are the coarsest bits and $r_{n_s-1},c_{n_s-1}$ are the finest bits. The $k$-th QTT site is assigned the pair $(r_k,c_k)$, and these two bits are stored on the physical qubits $(2k,2k+1)$. Thus, each qubit pair represents one spatial scale, and moving along the qubit register refines the pixel location from coarse to fine.
Given a computational basis state $i$, we read its bits as
\begin{equation}
b_{2k}(i)=r_k,
\quad
b_{2k+1}(i)=c_k,
\quad
k=0,\ldots,n_s-1 .
\label{eq:morton_layout}
\end{equation}
The corresponding pixel coordinate is decoded by the usual binary expansion,
\begin{equation}
r(i) = \sum_{k=0}^{n_s-1} r_k\,2^{n_s-1-k}, \quad
c(i) = \sum_{k=0}^{n_s-1} c_k\,2^{n_s-1-k}.
\label{eq:morton_decode}
\end{equation}
This defines a one-to-one mapping between computational basis states and pixels. Therefore, a Born probability $p_\theta(i\mid z)$ over basis states can be directly interpreted as a pixel distribution $p_\theta(r,c\mid z)$ under this fixed Morton mapping.

\paragraph{Alignment with the QTT sites.}
The layout matches the QTT structure: the site $A_k$ in \Cref{eq:qtt_classical} acts on the row-column bit pair $(r_k,c_k)$ that refines the spatial location at scale $k$. Routing coarse bits to lower qubit indices therefore aligns the qubit ordering with the QTT site ordering, so the circuit follows the same coarse-to-fine organization as the QTT decomposition.

\section{Site-block connectivity and gate counts}
\label{app:gate_counts}

This section specifies the entangler connectivity of the site blocks (\Cref{sec:qtt_skeleton}) and functions of the bond rank $\chi$, the number of scales $n_s$, and the per-site sub-layer count $L_{\text{site}}$. Throughout, $n_b = \log_2 \chi$ is the number of bond qubits, $n_{\text{tot}} = n_b + 2 n_s$ the total qubit count, and $W_{\text{site}} = n_b + 2$ the width of one local register.

\paragraph{Entangler connectivity.}
The CNOT pattern $E_{k,\ell}$ of site $k$ at sub-layer $\ell$ consists of
(i) a chain along consecutive bond qubits ($n_b - 1$ CNOTs);
(ii) for sites $k \geq 1$, two coarse-to-fine couplings from $(r_{k-1}, c_{k-1})$ to $(r_k, c_k)$, extending a chain along the physical qubits;
(iii) bond-to-physical couplings from every bond qubit to $r_k$ and to $c_k$ ($2 n_b$ CNOTs); and
(iv) one same-scale coupling $r_k \to c_k$.

\paragraph{Fused rotations.}
By the same-axis identity in \Cref{eq:angle_residual_identity}, the trainable and latent-injection rotations on a common axis can be fused into a single rotation per axis. The counts below assume this fused convention; an implementation that keeps the two rotations as separate gates doubles the site-block rotation count and leaves the CNOT count unchanged.

\paragraph{Single-qubit rotations.}
Each sub-layer applies one $R_y$ and one $R_z$ to every local qubit ($2 W_{\text{site}}$ rotations); there are $L_{\text{site}}$ sub-layers per site and $n_s$ sites, plus the opening layer $\prod_q R_y(\pi/2)$ on all qubits:
\begin{equation}
\#R_{1q} \;=\; n_{\text{tot}} \;+\; 2\, W_{\text{site}}\, L_{\text{site}}\, n_s .
\label{eq:rot_fused}
\end{equation}
The un-fused implementation instead uses
\begin{equation}
\#R^{\text{unfused}}_{1q} \;=\; n_{\text{tot}} \;+\; 4\, W_{\text{site}}\, L_{\text{site}}\, n_s .
\label{eq:rot_unfused}
\end{equation}

\paragraph{Two-qubit gates.}
Summing the pattern above, site $k$ contributes $3 n_b$ CNOTs per sub-layer for $k = 0$ and $3 n_b + 2$ for $k \geq 1$, so
\begin{equation}
\#\mathrm{CNOT} \;=\; L_{\text{site}}\, \big(\, 3\, n_b\, n_s + 2\,(n_s - 1)\,\big).
\label{eq:cnot_tot}
\end{equation}

\section{Latent modulation and data re-uploading}
\label{app:residual_vs_reuploading}

This section expands the relation, summarized in \Cref{sec:residual}, between latent modulation and the data re-uploading primitive it builds on \citep{perez2020data,schuld2021effect}.

\paragraph{Standard data re-uploading.}
Data re-uploading encodes a continuous input into rotation angles and injects it repeatedly across the circuit: at sub-layer $\ell$, qubit $q$, and axis $a$, a data-injection rotation $R_a(\omega^a_{\ell,q}\!\cdot z + \beta^a_{\ell,q})$ is placed next to the trainable rotation $R_a(\theta^a_{\ell,q})$. Its central theoretical property is the Fourier characterization: because $z$ enters the circuit repeatedly, the map $z \mapsto p_\theta(\cdot \mid z)$ is a finite trigonometric series whose frequency support is determined by integer combinations of the weights $\omega$. The primitive specifies repeated input injection, but leaves open how the trainable and data-injection rotations are parameterized and implemented.

\paragraph{Latent modulation.}
Latent modulation fixes this choice by using the same-axis composition identity. Since rotations about the same axis compose additively, the trainable and data-injection rotations fuse exactly into a single rotation, \Cref{eq:angle_residual_identity}, whose effective angle in \Cref{eq:effective_angle} is the sum of a trainable main path $\theta^a_{\ell,q}$ and an additive latent term $\beta^a_{\ell,q} + \omega^a_{\ell,q}\!\cdot z$. Two consequences follow. First, the rewriting is an algebraic identity rather than an approximation: relative to the unfused same-axis re-uploading form, it does not change the represented function class, and setting $\omega \equiv 0$ recovers the trainable rotation block up to the learnable offset $\beta$. Second, the fused form realizes the latent injection at a lower executed rotation count, quantified in \Cref{app:gate_counts}. Both formulations represent the same band-limited function class in $z$; what changes is the explicitness of the additive conditioning and the deployed gate cost.

\begin{table*}[htbp]
\centering
\small
\begin{tabular}{p{7cm}ccccc}
\toprule
Method families & Image-scale & Latent cond. & Decoder-free & Angle enc. & Structured prior \\
\midrule
Born machine \citep{liu2018differentiable}            & $\times$    & $\times$    & \checkmark & \checkmark & $\times$ \\
Quantum GAN \citep{dallaire2018quantum,lloyd2018quantum}                    & $\times$    & \checkmark  & \checkmark & \checkmark & $\times$ \\
Amplitude-encoded \citep{cacioppo2023quantum}           & $\times$    & \checkmark  & \checkmark & $\times$   & $\times$ \\
Hybrid quantum-classical \citep{huang2021experimental,chen2025quantum,islam2026quantum}      & \checkmark  & \checkmark  & $\times$   & \checkmark & $\times$ \\
SQGen (ours)                                    & \checkmark  & \checkmark  & \checkmark & \checkmark & \checkmark \\
\bottomrule
\end{tabular}
\caption{Comparison of representative quantum image-generation families. Columns: Image-scale, demonstrated on image-scale targets rather than only low-dimensional distributions; Latent cond., continuous-latent conditional generation; Decoder-free, no classical decoder in inference; Angle enc., angle-based latent encoding without amplitude state preparation; Structured prior, whether the deployed quantum generator itself incorporates an explicit problem-aligned structural prior.}
\label{tab:capability}
\end{table*}

\section{Torus latent prior versus a Gaussian prior}
\label{app:torus_prior}

\paragraph{Motivation.}
The latent $z$ enters the circuit through rotation angles \Cref{eq:site_sublayer}, and rotation angles are $2\pi$-periodic: $R_a(\vartheta)$ and $R_a(\vartheta+2\pi)$ implement the same gate. The prior on $z$ should therefore match the domain on which $z$ acts as a circuit input. We use the torus $\mathbb{T}^d \simeq [-\pi,\pi)^d$, with periodic identification of the boundary.

\paragraph{Comparison with a Gaussian prior.}
Classical latent-variable generators commonly use an isotropic Gaussian prior on $\mathbb{R}^d$ \citep{kingma2013auto}. Under angle encoding, however, a Gaussian prior is many-to-one: latents $\vartheta$ and $\vartheta+2\pi k$, for $k\in\mathbb{Z}^d$, induce the same rotations and therefore the same circuit. Thus, separated regions of Gaussian prior mass can alias to identical quantum inputs. Defining the latent directly on $\mathbb{T}^d$ removes this mismatch by making the latent support coincide with the periodic angle domain.

The encoder is matched to the torus prior through the wrap-respecting embedding $\Phi(z)=[\cos z,\sin z]$ in \Cref{eq:mmd_prior}. For sampling, we replace i.i.d.\ random draws with a scrambled Sobol quasi-Monte Carlo (QMC) point set on the torus. Sobol points provide more uniform coverage of the latent domain than independent random samples; after fixing the scrambling seed, they also give a reproducible deployment-time latent grid. In our setting, this makes the prior-matching objective more stable and ensures that deployment samples cover the periodic latent domain more evenly.

\section{Additional related work}
\label{app:related_extended}

Deep parameterized circuits are limited by barren plateaus, in which gradient variance vanishes exponentially with qubit count and depth \citep{mcclean2018barren,cerezo2021cost}, and hardware noise further flattens the optimization landscape on real devices \citep{wang2021noise}. Training Born machines through expectation-style objectives adds a measurement overhead that grows with circuit depth and gradient variance \citep{cerezo2021variational}. In addition, \citet{frkatovic2026generalization} present a detailed evaluation of quantum generative adversarial networks, reporting that images generated with a 16-qubit circuit and amplitude encoding collapse into noise-like output. \citet{lin2026structured} show that, starting from $16\times16$ images, amplitude-encoding circuits produce almost entirely noise due to deep preparation circuits. Together, these results underscore the challenges of current quantum image generation and motivate the design of SQGen: fully classical training with exact gradients and a structured circuit for quantum image generation.

\paragraph{Capability comparison.}
\Cref{tab:capability} summarizes representative families of quantum image generators along the design axes discussed in the related work section.

\begin{algorithm}[htbp]
\caption{SQGen training procedure.}
\label{alg:train_loop}
\begin{algorithmic}[1]
\STATE Sample minibatch $\{x_i\}_{i=1}^{B}$ and draw Sobol prior points $\{\tilde z_i\}_{i=1}^{B}\subset\mathbb{T}^d$.
\STATE Encode latents $z_i \gets E_\phi(x_i)$ and compute $p_i \gets p_\theta(\cdot\mid z_i)$ by batched differentiable tensor contraction.
\STATE Compute the reconstruction loss
\[
\begin{gathered}
\mathcal{L}_{\mathrm{rec}} = \frac{1}{B}\sum_{i=1}^{B} \mathcal{L}_{\mathrm{rec}}(p_i,q_{x_i}), \\
\mathcal{L}_{\text{rec}}(p_\theta,q_x) = \lambda_{\text{KL}}\, \mathrm{KL}(q_x\|p_\theta) + \lambda_{L_1}\, \|p_\theta-q_x\|_1.
\end{gathered}
\]
\STATE Compute the prior-matching loss
\[
\mathcal{L}_{\mathrm{prior}} = \mathrm{MMD}^2\!\left(\Phi(\{z_i\}_{i=1}^{B}), \Phi(\{\tilde z_i\}_{i=1}^{B})\right).
\]
\STATE Assemble the training objective
\[
\mathcal{L} = \mathcal{L}_{\mathrm{rec}} + \lambda_{\mathrm{prior}}\mathcal{L}_{\mathrm{prior}}.
\]
\STATE Backpropagate and take an AdamW step.
\end{algorithmic}
\end{algorithm}

\section{Training and evaluation details}
\label{app:setup_details}

This section expands the setup summarized in \Cref{sec:setup}.

\paragraph{Training.}
All configurations are optimized with AdamW \citep{loshchilov2017decoupled}, using QTT model learning rate $\eta_{\text{q}} = 2\!\times\!10^{-3}$, encoder learning rate $\eta_{\phi} = 1\!\times\!10^{-3}$, gradient clipping at norm $1.0$, and a differentiable forward pass. The batch size is $512$ for the plain backend and $256$ for the bond backend, and each model is trained for $30$ epochs of $50$ steps. \Cref{alg:train_loop} summarizes one training step. The $\lambda_{\text{KL}}$ is 0.5, $\lambda_{L_1}$ is 1.0, and $\lambda_{\text{prior}}$ is 2.0.

\paragraph{Latent and deployment.}
The latent dimension is $d = 4$ and the prior is a scrambled Sobol point set on $\mathbb{T}^4$. After training, the tensors are exported to a native-gate Qiskit circuit whose statevector output agrees with the differentiable QTT model.

\paragraph{Metrics.}
For each image $x$, the target $q_x$ is the image normalized to a 1024 distribution under the Morton ordering, and the model output is $p_x = p_\theta(\cdot \mid E_\phi(x))$. We report:
\begin{itemize}
\item \emph{Per-sample reconstruction}: $\mathrm{KL/x} = \mathbb{E}_x[\mathrm{KL}(q_x \,\|\, p_x)]$ and $\mathrm{L1/x} = \mathbb{E}_x[\|p_x - q_x\|_1]$.
\item \emph{Class-mean fidelity}: $\mathrm{KL}_{\text{c}} = \mathrm{KL}(\bar q \,\|\, \mathbb{E}_x[p_x])$ and $\mathrm{L1}_{\text{c}} = \|\mathbb{E}_x[p_x] - \bar q\|_1$ against the empirical class mean $\bar q$.
\item \emph{Diversity}: mean pairwise $L_1$ distance between sampled output distributions.
\end{itemize}

\begin{table}[t]
\centering
\setlength{\tabcolsep}{9pt}
\begin{tabular}{ccccc}
\toprule
Digit & $\mathrm{KL/x}$ & $\mathrm{L1/x}$ & $\mathrm{KL}_{\text{c}}$ & $\mathrm{L1}_{\text{c}}$ \\
\midrule
0 & 0.755 & 0.852 & 0.159 & 0.342 \\
1 & 0.818 & 0.849 & 0.350 & 0.505 \\
2 & 0.909 & 0.924 & 0.162 & 0.329 \\
3 & 0.830 & 0.883 & 0.177 & 0.353 \\
4 & 0.940 & 0.974 & 0.143 & 0.317 \\
5 & 0.930 & 1.024 & 0.183 & 0.373 \\
6 & 0.733 & 0.857 & 0.143 & 0.308 \\
7 & 0.868 & 0.933 & 0.120 & 0.311 \\
8 & 0.739 & 0.888 & 0.154 & 0.329 \\
9 & 0.754 & 0.870 & 0.117 & 0.315 \\
\midrule
Mean & 0.828 & 0.905 & 0.171 & 0.348 \\
$\pm$ std & $\pm 0.081$ & $\pm 0.058$ & $\pm 0.066$ & $\pm 0.059$ \\
\bottomrule
\end{tabular}
\caption{Per-digit MNIST metrics for SQGen.}
\label{tab:mnist_per_digit}
\end{table}
\begin{table}[t]
\centering
\small
\setlength{\tabcolsep}{4.5pt}
\begin{tabular}{lrrrrr}
\toprule
Method
  & $\mathrm{KL/x}$ $\downarrow$
  & $\mathrm{L1/x}$ $\downarrow$
  & $\mathrm{KL}_{\text{c}}$
  & $\mathrm{L1}_{\text{c}}$
  & Div.\ $\uparrow$ \\
\midrule
\multicolumn{6}{l}{\emph{Four configurations}} \\
Vanilla BM                & 0.399 & 0.561 & 0.070 & 0.177 & 0 \\
                          & $\pm 0.215$ & $\pm 0.204$ & $\pm 0.033$ & $\pm 0.049$ & / \\
BM + LM                    & 0.446 & 0.593 & 0.169 & 0.317 & \textbf{0.199} \\
                          & $\pm 0.255$ & $\pm 0.219$ & $\pm 0.155$ & $\pm 0.203$ & $\pm 0.161$ \\
BM + QTT                    & 0.428 & 0.586 & 0.100 & 0.221 & 0 \\
                          & $\pm 0.205$ & $\pm 0.191$ & $\pm 0.020$ & $\pm 0.026$ & / \\
SQGen & \textbf{0.394} & \textbf{0.550} & 0.121       & 0.271       & 0.175 \\
                          & $\pm 0.165$    & $\pm 0.166$    & $\pm 0.036$ & $\pm 0.071$ & $\pm 0.159$ \\
\midrule
Class mean      & 0.330 & 0.503 & 0     & 0     & 0 \\
Uniform         & 1.158 & 1.297 & 0.825 & 1.088 & 0 \\
\bottomrule
\end{tabular}
\caption{Fashion-MNIST image generation, aggregated over $10$ classes (mean $\pm$ std). The class-mean baseline trivially achieves $\mathrm{KL}_{\text{c}} = 0$ and is reported only as a lower-bound reference.}
\label{tab:fashion_aggregated}
\end{table}

\paragraph{Quantum hardware.}
\label{app:hardware}
The real-hardware experiments of \Cref{sec:real_hardware} run on the IBM Quantum Platform; the comparison in \Cref{fig:real} uses \texttt{ibm\_kawasaki} (Heron r2). The backend has 156 qubits and 176 couplers. 
For the results obtained, the reported median two-qubit (2Q) error was $1.56\times 10^{-3}$, the layered 2Q error was $5.32\times 10^{-3}$, and the best 2Q error was $6.48\times 10^{-4}$. The reported median readout error was $5.98\times 10^{-3}$. The median coherence times were $T_1=280.64\,\mu\mathrm{s}$ and $T_2=163.75\,\mu\mathrm{s}$. These calibration values provide context for the real-hardware results.

\section{Additional experiments and discussion}
This section reports additional experimental results, including per-digit MNIST metrics, Fashion-MNIST results, and a bond-rank sweep study.

\subsection{Per-digit MNIST metrics}
\label{app:mnist_per_digit}

\Cref{tab:mnist_per_digit} breaks the SQGen row of \Cref{tab:mnist_aggregated} down by digit. Digit $4$ is the hardest and digit $6$ the easiest by $\mathrm{KL/x}$. Digit $1$ has the largest class-mean gap, consistent with its concentrated class mean, for which any per-sample variation incurs the steepest class-mean penalty.

\subsection{Fashion-MNIST results}
\label{app:fashion}

\begin{table}[ht]
\centering
\setlength{\tabcolsep}{7pt}
\begin{tabular}{ccccc}
\toprule
Class & $\mathrm{KL/x}$ & $\mathrm{L1/x}$ & $\mathrm{KL}_{\text{c}}$ & $\mathrm{L1}_{\text{c}}$ \\
\midrule
T-shirt/top & 0.298 & 0.433 & 0.113 & 0.236 \\
Trouser     & 0.335 & 0.502 & 0.123 & 0.292 \\
Pullover    & 0.289 & 0.470 & 0.105 & 0.243 \\
Dress       & 0.315 & 0.460 & 0.068 & 0.191 \\
Coat        & 0.305 & 0.435 & 0.126 & 0.236 \\
Sandal      & 0.829 & 0.984 & 0.197 & 0.422 \\
Shirt       & 0.314 & 0.469 & 0.103 & 0.222 \\
Sneaker     & 0.350 & 0.555 & 0.098 & 0.242 \\
Bag         & 0.478 & 0.643 & 0.162 & 0.369 \\
Ankle boot  & 0.432 & 0.553 & 0.121 & 0.254 \\
\midrule
Mean & 0.394 & 0.550 & 0.121 & 0.271 \\
$\pm$ std & $\pm 0.165$ & $\pm 0.166$ & $\pm 0.036$ & $\pm 0.071$ \\
\bottomrule
\end{tabular}
\caption{Per-class Fashion-MNIST metrics for SQGen.}
\label{tab:fashion_per_class}
\end{table}

\Cref{tab:fashion_aggregated} repeats the four-configuration comparison of \Cref{sec:main_results} on Fashion-MNIST. The architecture and evaluation protocol are unchanged from MNIST. SQGen attains the best per-sample reconstruction of the four configurations. As expected: SQGen is again the only configuration that combines low per-sample error with non-zero diversity. \Cref{tab:fashion_per_class} gives the per-class breakdown of SQGen metrics.

\subsection{Bond-rank sweep aggregated over MNIST digits}
\label{app:bondrank_agg}

\begin{table}[htbp]
\centering
\small
\setlength{\tabcolsep}{3pt}
\begin{tabular}{lcccc}
\toprule
Method & $\chi$ & $\mathrm{KL/x}$ & $\mathrm{L1/x}$ & Div.\\
\midrule
BM+QTT & 2  & $0.908 \pm 0.086$ & $0.982 \pm 0.065$ & $0$ \\
BM+QTT & 4  & $0.876 \pm 0.080$ & $0.964 \pm 0.067$ & $0$ \\
BM+QTT & 8  & $0.879 \pm 0.078$ & $0.970 \pm 0.063$ & $0$ \\
BM+QTT & 16 & $0.897 \pm 0.089$ & $0.979 \pm 0.064$ & $0$ \\
\midrule
SQGen & 2  & $0.927 \pm 0.125$ & $0.962 \pm 0.079$ & $0.289 \pm 0.177$ \\
SQGen & 4  & $0.942 \pm 0.235$ & $0.987 \pm 0.144$ & $0.311 \pm 0.236$ \\
SQGen & 8  & $0.828 \pm 0.081$ & $0.905 \pm 0.058$ & $0.214 \pm 0.175$ \\
SQGen & 16 & $0.923 \pm 0.143$ & $0.969 \pm 0.091$ & $0.115 \pm 0.102$ \\
\bottomrule
\end{tabular}
\caption{Bond-rank sweep aggregated over the ten MNIST digits (mean $\pm$ std across digits), $\chi \in \{2, 4, 8, 16\}$.}
\label{tab:bondrank_agg}
\end{table}

\Cref{tab:bondrank_agg} reports the bond-rank sweep of \Cref{sec:ablations} averaged over the ten MNIST digits, complementing the digit-$0$ view in \Cref{tab:bondrank}. Mean reconstruction is best at $\chi = 8$, while diversity is highest at small $\chi$ and decreases as the bond rank grows; this trade-off underlies the default choice $\chi = 8$ discussed in \Cref{sec:ablations}.

\section{Depolarizing-noise model}
\label{app:depolarizing_noise}

The noise study in \Cref{sec:noise} uses a two-qubit depolarizing channel applied after every CNOT gate. This is a synthetic circuit-noise model; its purpose is to isolate how the circuit output changes as the strength of a simple per-entangling-gate perturbation increases.

$\rho$ denote the density matrix before a CNOT gate with ideal unitary $U_{\mathrm{CNOT}}$ on two qubits. The ideal update $\rho' = U_{\mathrm{CNOT}}\, \rho\, U_{\mathrm{CNOT}}^\dagger$ is followed by the depolarizing channel \citep{nielsen2010quantum}
\begin{equation}
\mathcal{D}_p(\rho') \;=\; \Big(1-\tfrac{15p}{16}\Big)\rho' \;+\; \frac{p}{16} \sum_{P \in \mathcal{P}_2 \setminus \{I \otimes I\}} P \rho' P^\dagger ,
\end{equation}
where $\mathcal{P}_2 = \{I, X, Y, Z\}^{\otimes 2}$ is the set of two-qubit Pauli operators and $p \in [0,1]$ is the depolarizing rate.

The channel is inserted after every CNOT gate and not after single-qubit rotations, so circuits with more two-qubit gates accumulate more perturbations at the same $p$. Since the bond circuits contain more CNOTs as $\chi$ increases (\Cref{app:gate_counts}), this model directly probes the trade-off between cross-scale capacity and sensitivity to two-qubit noise. We evaluate the noisy circuits with exact density-matrix simulation (Qiskit Aer) and read the output probabilities from the diagonal of the final density matrix.

\end{document}